# The impact of intrinsic magnetic field on the absorption signatures of elements probing the upper atmosphere of HD209458b

M. L. Khodachenko[1,3,5], I. F. Shaikhislamov[2,3,4], H. Lammer[1],
I. B. Miroshnichenko[2,4], M. S. Rumenskikh[2,3,4], A. G. Berezutsky[2,3], L. Fossati[1]

*1) Space Research Institute, Austrian Academy of Sciences, Graz, Austria*
*2) Institute of Laser Physics SB RAS, Novosibirsk, Russia*
*3) Institute of Astronomy, Russian Academy of Sciences, Moscow, Russia*
*4) Novosibirsk State Technical University, Novosibirsk, Russia*
*5) Lomonosov Moscow State University, Skobeltsyn Institute of Nuclear Physics, Moscow, Russia*
*E-mail address: maxim.khodachenko@oeaw.ac.at; ildars@ngs.ru*

**ABSTRACT**

The signs of an expanding atmosphere of HD209458b have been observed with far-ultraviolet transmission spectroscopy and in the measurements of transit absorption by metastable HeI. These observations are interpreted using the hydrodynamic and Monte-Carlo numerical simulations of various degree of complexity and consistency. At the same time, no attempt has been made to model atmospheric escape of a magnetized HD209458b, to see how the planetary magnetic field might affect the measured transit absorption lines. This paper presents the global 3D MHD self-consistent simulations of the expanding upper atmosphere of HD209458b interacting with the stellar wind, and models the observed HI (Lyα), OI (1306 Å), CII (1337 Å), and HeI (10830 Å) transit absorption features. We find that the planetary dipole magnetic field with the equatorial surface value of $B_p = 1$ G profoundly changes the character of atmospheric material outflow and the related absorption. We also investigate the formation of planetary magnetosphere in the stellar wind and show that its size is more determined by the escaping atmosphere flow rather than by the strength of magnetic field. Fitting of the simulation results to observations enables constraining the stellar XUV flux and He abundance at $F_{XUV} \sim 10$ erg cm$^2$ s$^{-1}$ (at 1 a.u.) and He/H≈0.02, respectively, as well as setting an upper limit for the dipole magnetic field of $B_p < 0.1$ G on the planetary surface at the equator. This implies that the magnetic dipole moment $\mu_P$ of HD209458b should be less than 6% of the Jovian value.

**1. INTRODUCTION**

The study of exoplanets is a rapidly developing branch of space science. About 8% of confirmed exoplanets (380 of 4375) belong to the group of "hot Jupiters" (hereafter HJs) and "warm Neptunes", i.e. have the mass of (0.05-5)$M_{Jup}$, radius of (0.2-2)$R_{Jup}$, and orbit semi-major axes <0.2 a.u. (see e.g. in https://exoplanetarchive.ipac.caltech.edu/index.html ). Many of them (221) are located extremely close to their host stars, at the distances shorter than 0.05 a.u.[1]. This leads to photo-ionization, heating, and hydrodynamic escape of the upper atmospheres of these planets. One of the most studied and widely discussed is HD209458b. *Vidal-Madjar et al.* (*2003*) observed the primary transits of this planet in Lyα with the STIS spectrograph on board of HST and reported 15% absorption in the high velocity blue wing of the line. The re-analyses of the same data by *Ben-Jaffel (2007), Ben-Jaffel & Sona Hosseini* (*2010*) and *Vidal-Madjar et al. (2008)* gave a bit smaller (6-9%) and more symmetric absorption profiles. The primary transit measurements of HD209458b at far UV wavelengths (with HST/STIS) also revealed the absorption depths of 10±4.5% in OI ($2p^2$ 2P–$2p^2$ 2D) and 7.5±3.5% in CII ($2p^4$ 3P–$2p^4$ 3S) resonance lines (*Vidal-Madjar et al. 2004*). Additional analysis of this data-set by *Ben-Jaffel & Sona Hosseini* (*2010*) led to similar results (10.5±4.4% and 7.4±4.7% for OI and CII, respectively). *Linsky et al.* (*2010*) analyzed four additional far UV transit observations by COS/HST spectrograph and confirmed the depth of 7.8±1.3% in CII line. *Linsky et al. (2010)* reported also the detection of absorption in SiIII (1206.5 Å) resonance line at a level of 8.2±1.4%, for which V*idal-Madjar et al. (2004)* earlier reported the non-detection. The latest up to date UV observations of HD209458b carried out with HST/STIS and their analysis have shown the presence of FeII in the upper atmosphere (*Cubillos et al. 2020*) as well as 6.2±2.9% absorption in the MgI resonance line (*Vidal-Madjar et al. 2013*), while no absorption was detected in the MgII h&k lines. Altogether, magnesium probably might be locked in aerosols in the lower atmosphere of the planet. Anyway, all these data point at the presence of heavy elements in the expanding upper atmosphere of HD209458b, which escapes the planet beyond the Roche lobe.

---
[1] The numbers in brackets reflect the status of Apr. 2021

A new opportunity to probe the expanding atmospheres of HJs and warm Neptunes is opened by the measurements of the metastable helium $2^3S$ triplet line at 10830 Å (hereafter HeI($2^3S$); *Seager & Sasselov 2000; Oklopčić & Hirata 2018*). For HD209458b the transit at 10830 Å was measured by *Alonso-Floriano et al. (2019)* using the ground-based Calar Alto CARMENES instrument. The detected excess absorption at the line center was at the level of 0.91±0.10% with the width of the line 11 km/s in Doppler shifted velocity units. The $9\sigma$ fidelity level leaves no doubt regarding the presence of HeI in the upper atmosphere of HD209458b, which occupies a region of at least 2-3 planet radii $R_p$ around the planet. Moreover, according to the recent estimates, the high resolution spectropolarimetric transit observations at the wavelength of HeI($2^3S$) 10830 Å line could be used for the detection of exoplanetary magnetic fields (*Oklopčić et al. 2020*) in a similar way as done during the exploration of magnetic fields in solar coronal filaments.

Such an extensive observational material available for the HD209458b, makes this HJ a primary candidate for the application of complex numerical models aimed at interpretation of the measurements and inferring physical parameters and features of the stellar-planetary system, such as the mass loss of the planet, abundances of its upper atmospheric species, the parameters of stellar wind (hereafter SW) flow, and the stellar radiation flux. Such models have been continuously developed over the last decade by different research groups, while increasing in complexity and details of the simulated phenomena. In particular, the energy limited approach applied by *Lammer et al. (2003)* revealed that the hydrogen-dominated atmospheres of HJs are heated by the stellar XUV radiation up to several thousand Kelvin and driven to continuous expansion and escape. Soon after discovery of first HJs, several 1D hydrodynamic (HD) models have been developed (*Yelle 2004; García Muñoz 2007; Penz et al. 2008; Murray-Clay et al. 2009; Guo 2011; Koskinen et al. 2010, 2013; Shaikhislamov at al. 2014; Salz et al. 2016,* and references therein) to simulate their upper atmospheres. These models have shown that the upper atmosphere heating and ionization by the stellar XUV radiation leads to an outflow of planetary gas. It forms an escaping planetary wind (hereafter PW), that overcomes the planet gravity and expands beyond the Roche lobe with a supersonic velocity.

The first quantitative interpretation of the Lyα transit observations of HD209458b, proposed in *Vidal-Madjar et al. (2003)* and *Lecavelier des Etangs et al. (2004, 2008),* assumed the acceleration of planetary hydrogen atoms by the stellar Lyα radiation pressure, which according to the estimations made for an isolated atom produces a significant force *(Bourrier & Lecavelier des Etangs 2013)*. This acceleration and the resonant Doppler broadening mechanism were invoked to explain the 6-9% absorption at the blue wing of the Lyα line, measured for HD209458b. At the same time, the 1D HD modeling has shown that the density of atomic hydrogen in the Roche lobe of the planet is high enough to cause the absorption over the whole emission profile due to the natural line broadening mechanism (*Garcia Muñoz 2007, Ben-Jaffel 2007, 2008, Koskinen et al. 2010*). This conclusion was supported in *Shaikhislamov et al. (2016), Khodachenko et al. (2017)* and *Shaikhislamov et al. (2018a)*, who used a 2D multi-fluid simulations and treated self-consistently the interaction of PW with the flow of SW plasma. The 1D models were also used to simulate and interpret the measured absorption by metastable HeI($2^3S$) atoms for the HD209458b (*Alonso-Floriano et al. 2019, Lampón et al. 2020*), inferring the helium abundance in the upper atmosphere in the range of 0.02≤He/H<0.1.

However, the 1D HD models have difficulties with explanation of the observed absorption depths in OI, CII and SiIII lines for HD209458b at the levels comparable to (or even exceeding) those measured in Lyα line. For example, *Ben-Jaffel & Sona Hosseini* (*2010*), using the density distributions from *Garcia Muñoz (2007)*, calculated for the OI and CII multiplets the absorption of only 3.9% and 3.3%, respectively. Using a 1D hydrostatic model, *Koskinen et al. (2010)* calculated the absorption by OI atoms at the level of 4.3%, whereas for CII and SiIII the obtained values were 4.6% and 5.2%, respectively (*Koskinen et al. 2013*). As summarized in *Shaikhislamov et al. (2018a)*, while the Lyα absorption can be explained by the natural line broadening with a sufficiently large number of hydrogen atoms contained inside the Roche lobe (*Khodachenko et al. 2017*), the density of heavier species is too small for that, whereas the typical temperature of $10^4$ K is too low for the producing of significant resonant absorption by the heavy particles over the typical line half-widths of 20–30 km/s. The only reasonable mechanism to produce the line broadenings comparable to the observations, is related with a global flow of the escaping planetary atmospheric material outside the Roche lobe. To simulate the PW motion beyond the Roche lobe, at least a 2D HD model is needed. The calculation of transit absorption for HD209458b with the 2D and 3D gasdynamic codes was done for the first time in our previous papers (*Shaikhislamov et al. 2018a, Shaikhislamov et al. 2020a*). It was shown that the escaping PW of HD209458b is sufficiently fast to generate the observed significant Doppler resonant absorption in the lines of OI and CII, assuming their solar abundances.

Planetary intrinsic Magnetic Field (hereafter MF) is another important factor which affects the PW and SW interaction and the related transit absorption features. Theoretical estimates (*Grießmeier et al. 2004, 2005, 2007, Khodachenko et al. 2012*) predict that the tidally locked gas giants should have significantly smaller intrinsic MFs than those of their Solar System's analogues. There is also an alternative point of view (*Reiners & Christensen, 2010*), based on the scaling properties of evolution of the convection-driven dynamos (*Christensen & Aubert 2006*), which predicts rather high intrinsic MFs of HJs, decreasing during the lifetime of a planet, as well as independence of the magnetic dipole moment $\mu_P$ on planet rotation (*Reiners & Christensen 2010*). Therefore the estimation of planetary intrinsic MF still remains an open issue. It is worth to mention in this respect that the numerical simulation of the MF-sensitive measurable phenomena, such as transit absorption of different spectral lines dependent of the planetary material dynamics affected by the MF, opens a way for indirect probing of the exoplanetary MF by fitting the numerical modelling results to observations. An example of such study is presented in this paper.

The first works, dedicated to the effects of MF on HJs, studied the problem in 1D (*Trammell 2011*), and later in 2D geometry (*Trammell et al. 2014, Owen 2014, Khodachenko et al. 2015*). They have revealed the presence of the so-called 'dead' and a 'wind' zones formed during the interaction of the escaping PW flow with the intrinsic dipole MF of the planet. Along with that, a number of complex 3D MHD simulations have been performed, using the already existing open source astrophysical codes, adopted to the problem of PW and SW interaction. These involved, however, a simplified description of the HJ's PW, as a boundary condition at the heights of the thermosphere temperature maximum (~$2R_p$), or a prescribed, already formed, supersonic PW flow at the heights of $(4 \div 6)R_p$. Moreover, instead of a complete aeronomy calculations to simulate the behavior of the expanding upper atmosphere of hot exoplanets, only the dynamics of planetary protons, and in some cases of hydrogen atoms, was considered. For example, using the public code PLUTO in an isothermal approximation ($\gamma=1.05$), *Matsakos et al. (2015)* investigated different regimes of interaction of a typical magnetic HJ with a magnetized SW. The calculations were performed on a Cartesian grid in the simulation domain, which includes the star and the planetary orbit (so-called global modeling). This approach enabled identifying several basic structures and PW-SW interaction phenomena, e.g., the bow-shock, tail, and accretion streams of planetary material, controlled by the planetary MF and the parameters of interacting material flows. Using the same code, *Daley-Yates & Stevens (2018, 2019)* estimated the generation of detectable radio emission at a HJ, as well as investigated the accretion of planetary atmospheric material on the star, showing strong dependence of these processes on the MF topology formed by the dipole fields of the planet and star. By combining 1D and 3D calculations, *Erkaev et al. (2017)* modeled the interaction of PW with a magnetized sub-Alfvénic SW. They studied the formation of magnetic compression layer, which significantly affects the production of hydrogen energetic neutral atoms (ENAs) and the related absorption in the Ly$\alpha$ line. At the same time, the interaction between PW and SW, addressed in this study, was not treated in a self-consistent way.

One of the advanced MHD models of a magnetic HJ interacting with the magnetized SW is described in (*Arakcheev et al. 2017, Zhilkin et al. 2020*). It is implemented on an irregular Cartesian grid, centered on the planet, while the SW is modeled within the simulation domain by applying a boundary condition, which is analytically derived from the Parker's solution (*Parker 1958*). This model enabled quantifying several effects relevant to the observations, e.g., suppression of PW by the planetary MF (*Arakcheev et al. 2017*) and the influence of the interplanetary MF on the structure of the escaping PW (*Zhilkin & Bisikalo 2019, Zhilkin et al. 2020*). However, this model is still based on a one-fluid MHD approach and, therefore, it lacks the aeronomy of the planetary upper atmosphere. This limitation is common for the majority of currently used MHD models. It restricts their applicability for the self-consistent simulation of transit spectra. An attempt to relate observations with the effects of planetary MF was made by *Kislyakova et al. (2014)*. Using a kinetic Monte-Carlo model and a parametrized planetary magnetospheric obstacle, they tried to reproduce and fit the measured for HD209458b Lyα transit absorption profiles, and estimated in this way the value of the planetary magnetic dipole moment, $\mu_P$, which appeared to be ~10% of that of the Jupiter. At the same time, the escaping PW and the magnetosphere itself were not modeled in this study, but prescribed, whereas the interaction of escaping planetary plasma flow with the background planetary MF was ignored.

Indeed, the realistic simulation of absorption in the lines of various species in presence of planetary MF requires a 3D multi-fluid MHD code, which self-consistently combines the aeronomy of upper atmosphere and the global dynamics of the SW. While the global 3D MHD models have been already used for the study of stellar-planetary interactions, the aeronomy codes for the simulation of hot exoplanetary atmospheres only recently began to be upgraded from 1D to 3D geometry (*Tripathy et al 2015, Shaikhilamov et al.2018b, Wang & Dai 2018*). In the present paper we make a step forward on the way of creation of a global multi-fluid 3D MHD self-consistent model of the interacting PW and SW of a HJ, and simulate with it the transit absorption features of HD209458b in the most of the observed spectral lines, such as HI(Lyα), OI, CII and HeI($2^3$S), taking into account the influence of a planetary MF. We exclude from the consideration the lines of Si and Mg, because their abundances in the upper atmosphere of HD209458b, due to the aerosol condensation at the low altitudes are too much unconstrained. It should be noted, that inclusion into consideration of the HeI($2^3$S) triplet quantitatively and qualitatively broadens the base for comparison between the observations and simulations. We do not describe here in details the specific planetary MF related phenomena, such as magnetodisk, magnetotail, and associated with them dynamic reconnection processes. To certain extend, they have been addressed generally in our previous papers in *Khodachenko et al. (2012, 2015)* based on more simplified modelling approaches. At the same time, given the increased capability of the global self-consistent multi-fluid MHD model, all these phenomena can be studied in more detail. This work, however, deserves a separate dedicated paper. Instead, we focus on the study of how these MF related phenomena, if really present, might affect the available observations.

The paper is organized as follows. In Section 2, we briefly describe our numerical model and its novel features. In Section 3 the simulation results on the absorption in different lines are presented, and their dependence on the SW and planetary MF conditions is investigated. Section 4 contains the discussion and conclusions.

## 2 THE NUMERICAL MODEL AND BASIC ISSUES

The global 3D multi-fluid HD model used as a basis in the present work has been already described in our earlier papers, e.g., *Shaikhislamov et al. (2018b, 2020a,b,c)* and *Khodachenko et al. (2019, 2021)*. It was developed as an upgrade of the previous 1D (*Shaikhislamov et al. 2014*) and 2D (*Khodachenko et al. 2015, 2017, Shaikhislamov et al. 2016*) models. The model code solves numerically the hydrodynamic equations of continuity, momentum, and energy for all species of the simulated multi-component PW and SW plasmas. Among the considered species, the model, used in this paper, includes hydrogen and helium particles (H, $H^+$, He, $He^+$, $He^{2+}$) as well as the heavier particles, O and C in atomic and up to Z=3 ionization levels. While the model in its general version includes also the molecular hydrogen species, such as $H_2$, $H_2^+$, and $H_3^+$ (*Shaikhislamov et al. 2018b, Khodachenko et al. 2019*), in the present study we do not take them into account. The reason for such simplification is due to the quick dissociation of these complex species at the relatively low altitudes, and therefore their insignificant influence on the overall PW dynamics, demonstrated for the case of HD209458b in our previous studies (*Shaikhislamov et al. 2014, 2016, Khodachenko et al. 2015, 2017*). Besides of that, as it was shown with 1D aeronomy simulation of HD209458b, performed by *García Muñoz (2007)*, the presence of oxygen with an abundance higher than just 1% of the Solar System value leads to a rapid decrease of the $H_2$ population with height. Without accounting for molecular species, the H, He, C, and O chemistry is governed by photo-ionization, dielectronic recombination, electron impact excitation and ionization. Photo-ionization results also in the strong heating of the gas by the produced photo-electrons, which drives the hydrodynamic outflow of the planetary upper atmospheric material. The corresponding heating term is derived by integration of the absorption cross-sections of species over the XUV spectrum (λ<912 Å), taking into account the attenuation of photons in the atmosphere according to the wavelength-dependent cross-section. As a proxy for the XUV flux, $F_{XUV}$, of the solar-like star HD209458, we use the solar spectrum. However, for the near-ultraviolet (NUV) (912<λ<3000 Å) and near-infrared (NIR) parts of the stellar radiation flux, $F_{NUV}$ and $F_{NIR}$, respectively we employ a synthetic stellar spectral energy distribution (SED), computed with the LLmodels stellar atmosphere code (*Shulyak et al. 2004*), specifically for HD209458. These are needed for the calculation of photo-ionization of the metastable HeI($2^3$S) atoms and the radiation pressure acting on them.

The major novelty of this work consists in the inclusion of planetary MF into the previously developed global multi-fluid HD model and its extension to an MHD one. The elaborated new multi-fluid MHD model self-consistently simulates the expansion and escape of the multi-species exoplanetary upper atmosphere, driven by the stellar radiative heating and gravitational forces, and its interaction with the surrounding SW plasma flow, which is also simulated within the model. This model extension is achieved by the addition to the set of HD equations of the MF induction equation, as well as the inclusion of the Ampere force in the momentum equations for the ionized species. In the most of the area around the planet, because of the large typical scales of the considered system, the dynamics of MF is assumed to be dissipation-less, which in the numerical model is achieved by taking a sufficiently high, though finite, electric conductivity σ, which corresponds to the magnetic Reynolds number $R_m \sim 10^5$. This value of the magnetic Reynolds number was found empirically to exceed the numerical diffusion in the MF induction equation. The set of MHD equations is solved in a global non-inertial spherical frame, centered at the planet and rotating in phase with its orbital motion, i.e., in a so-called tidally locked frame of reference. The polar-axis, Z, and the planetary magnetic dipole moment, $\mu_P$, are directed perpendicular to the equatorial plane, coplanar to the ecliptic. To keep the number of points in the numerical code tractable for processing, the applied radial mesh is highly non-uniform, with the grid step increasing linearly from the planet surface. This allows resolving of the highly stratified upper atmosphere of the planet, where the grid step is as small as $\Delta r = R_P/400$. The latitudinal mesh also varies, typically from $\Delta\theta=0.05$ at the equator up to $\Delta\theta=0.15$ at the polar axis to reduce refining of the azimuthal mesh at the poles. The numerical scheme is fully explicit, with the second order accuracy of the space and time discretization. We calculate the MF induction equation assuming that the divergence of the MF is zero, however we do not clean specially the divergence. At the inner boundary of the computation domain, i.e. at the provisional optical radius of the planet r=$R_P$, we fix the MF flux by fixing the radial component of the magnetic dipole field $B_r$=const, while the perturbations of the azimuthal and poloidal components of the field obey an open boundary condition $\partial_r(r \cdot \delta B_\perp) = 0$.

For the self-consistent calculations on the scale of the whole star-planet system, we incorporate in the same code the simulation of the SW plasma flow, parametrized with the total stellar mass loss rate, $M'_{SW}$, coronal temperature, $T_{cor}$, and terminal SW speed, $V_{SW,\infty}$ (see, e.g., in *Khodachenko et al. 2019, Shaikhislamov et al. 2020a*). For these parameters we take the typical solar values. As it was shown in our previous study of HD209458b (*Shaikhislamov et al. 2020a*), such SW does not produce much influence on the absorption in the considered lines, because the bow-shock and the whole PW-SW interaction region are located relatively far from the planet. So, they do not affect the dynamics of the expanding PW in the inner regions, close to the planet, where the absorption processes take place. Nevertheless, in a separate section in this paper we specially explore the case of a strong SW, which is ten times denser and 1.5 times faster, than the typical solar wind. Our model also admits the account of the radiation pressure acting on atoms, with the inclusion of the effect of self-shielding of photons. As it was shown in our previous papers (*Khodachenko et al. 2017, Shaikhislamov et al. 2020a*) and confirmed in *Debrecht et al. (2020)*, the radiation pressure produced by Lyα photons on hydrogen atoms is not significant under the conditions of HD209458b, so we do not include it in the present study. However, the NIR radiation pressure acting on the metastable HeI($2^3$S), which was shown to be an important factor in other exoplanetary systems (e.g., Wasp107b, *Allart et al. 2019, Khodachenko et al. 2021*), is properly taken into consideration.

The details regarding calculation of the HI(Lyα), OI and CII absorption are described in *Shaikhislamov et al. (2020a)*. The measured out of transit and the reconstructed Lyα line shapes of HD209458 are given in (*Vidal-Madjar et al. 2003*). The ±50 km/s interval of the Doppler velocities around the Lyα line center is excluded from consideration because of geocoronal contamination. The shape of the CII emission triplet was measured by *Linsky et al. (2010)*. For the OI triplet, the proxy of the line shapes are based on the solar spectra measurements. For the averaging of absorption over the triplets, the populations of upper sublevels of the absorbing elements are also taken into account according to their statistical weights and relative emission amplitudes. For the OI $2p^2$ 2P–$2p^2$ 2D and CII $2p^4$ 3P–$2p^4$ 3S transitions at 1302.2 Å and 1334.5 Å, respectively, which are strongly absorbed in the interstellar medium (ISM), we cut the line core regions of ±6 km/s, whereas for the other lines of the triplets, not affected by the ISM, the whole shapes are used. For the calculation of absorption at HeI($2^3$S) metastable triplet no line shapes are needed. It is not affected by the ISM and the upper sublevels are also populated according to their statistical weights (*Seager & Sasselov 2000, Oklopčić & Hirata, 2018*).

**3 SIMULATION RESULTS**

Further on, if not specified otherwise, we use the following 'standard' conditions for the simulations: the stellar NUV flux, $F_{NUV}=5·10^4$ erg cm$^{-2}$ s$^{-1}$; abundances, O/H=8.5·10$^{-4}$, C/H=3.6·10$^{-4}$; the inner boundary temperature at the position of conventional planetary surface, i.e. at $r=R_p$, $T_{base}$=1500 K. Besides of that, velocity, temperature and density of SW at the orbit of planet are $V_{sw}$=220 km/s, $T_{sw}$=8.5·10$^5$ K, and $n_{sw}$=4·10$^3$ cm$^{-3}$, respectively which are typical for a moderate SW. These correspond to the total stellar mass loss rate $M'_{SW}$=10$^{12}$ g/s and coronal temperature $T_{cor}$~ 2·10$^6$ K (see e.g. in *Khodachenko et al. 2019*). Altogether, this set of parameters forms a basis for the most of the simulation runs, which differ one from another by other modelling parameters specified below (see Table 1).

During the reported here study of the possible effects of planetary MF on the dynamics of the escaping PW of HD209458b and related transit absorption features, we varied the XUV flux in the range $F_{XUV}$=(4÷15) erg cm$^{-2}$ s$^{-1}$, helium abundance in the range He/H = 0.05÷0.1, as well as considered the cases of magnetized and non-magnetized planet. Here and further on, the radiation flux values are given for the reference distance of 1 a.u. We note that the parameters of the model run N1 are close to those used in our previous studies of HD209458b with the 2D and 3D HD models (*Shaikhislamov et al. 2018a, 2020a*). At the inner boundary of the simulation domain, we assume a pure atomic atmosphere, taking into account the fast dissociation of molecular hydrogen mentioned in the previous section. The typical 3D picture of the PW flow immersed in the SW at such conditions has been described in details in *Shaikhislamov et al.* (*2020a*).

In the next subsections we present the results of our simulations of the transit absorption features of HD209458b under different conditions and the corresponding parameters of the model, which all are summarized in Table 1.

*3.1 Absorption at 10830 Å*

The main novelty of the present study, besides of the account of possible planetary MF, consists in the calculation of synthetic transit absorption by the HeI($2^3$S) atoms and comparing it with the observations. The crucial difference of HD209458b, as compared to other exoplanetary systems we have studied with respect to absorption at HeI($2^3$S) 10830 Å line, is that the NUV flux of the host HD209458 star appears to be very strong due to the relatively large stellar size and high temperature of photosphere. Because of that, the main depopulation process of the metastable HeI($2^3$S) is the photo-ionization (with a threshold at 2600 Å). For example, if in the cases of GJ3470b and Wasp107b the photo-ionization time of HeI($2^3$S) is ~17 min and ~12 min, respectively, for the HD209458b it is only 0.3 s. This time is sufficiently short to reduce the effect of the radiation pressure on HeI($2^3$S) atoms, which otherwise would accelerate them up to extremely high velocities. Indeed, for the NIR (λ=10830 Å) flux of $F_{10830}$ ~ 90 erg cm$^{-2}$ s$^{-1}$ Å$^{-1}$ at 1 a.u. the radiation pressure force acting on HeI($2^3$S) atoms is 460 times higher than that of the stellar gravity, and the corresponding acceleration of atoms in the region of the planet orbit reaches the values of ~1.5 km/s$^2$. The HeI($2^3$S) atoms are created mainly by the recombination of He$^+$ ions, which in turn are produced by the photoionization of HeI($1^1$S) by the high energy XUV radiation at λ<512 Å. Therefore, the HeI($2^3$S) absorption at 10830 Å depends on the XUV flux $F_{XUV}$ and helium abundance. Both these parameters are in fact poorly constrained. At the same time, the NIR flux, $F_{NIR}$, and therefore also $F_{10830}$, provided by the employed synthetic stellar SED, are fixed for the given parameters of the host star, HD209458. So, they cannot be varied in the simulations, whereas the range of possible variation of the NUV flux, $F_{NUV}$, due to the uncertainty on the stellar parameters can be only within a factor of 2.

| N | $F_{XUV}$ [erg cm$^{-2}$ s$^{-1}$], at 1 a.u. | $F_{NUV}$ [erg cm$^{-2}$ s$^{-1}$], at 1 a.u. | He/H | $M'_p$ [×10$^{10}$g/s] | $A_{HI}$ % | $A_{OI}$ % | $A_{CII}$ % | $A_{HeI}$ % | $B_p$ [G] / $\mu_P$[$\mu_J$] | $M'_{sw}$ [×10$^{10}$g/s] |
|---|---|---|---|---|---|---|---|---|---|---|
| 1 | **7.5** | 5·10$^4$ | **0.05** | 16.9 | 8.0 | 3.7 | 3.2 | 0.65 | 0 / 0 | 10$^2$ |
| 2 | **4.0** | 5·10$^4$ | **0.1** | 7.4 | 7.3 | 3.2 | 2.7 | 0.67 | 0 / 0 | 10$^2$ |
| 3 | **15** | 5·10$^4$ | 0.02 | 32.4 | 9.0 | 6.3 | 5.9 | 0.45 | 0 / 0 | 10$^2$ |
| 4 | **10** | 3·10$^4$ | 0.02 | 25.8 | 9.1 | 6.5 | 6.1 | 0.51 | 0 / 0 | 10$^2$ |
| 5 | **4.0** | 2.5·10$^4$ | 0.02 | 13.4 | 12 | 4.4 | 3.7 | 0.28 | 0 / 0 | 10$^2$ |
| 6 | 10 | 3·10$^4$ | 0.02 | 11.8 | 6.7 | 4.0 | 3.7 | 0.48 | **1 / 0.61** | 10$^2$ |
| 7 | 10 | 3·10$^4$ | 0.02 | 16.8 | 7.6 | 4.3 | 4.3 | 0.40 | **0.5 / 0,3** | 10$^2$ |
| 8 | 10 | 3·10$^4$ | 0.02 | 20.4 | 8.6 | 5.3 | 5.1 | 0.41 | **0.25 / 0.15** | 10$^2$ |
| 9 | 10 | 3·10$^4$ | 0.02 | 23.1 | 8.8 | 6.0 | 5.7 | 0.45 | **0.1 / 0.061** | 10$^2$ |
| 10 | 10 | 3·10$^4$ | 0.02 | 25.8 | 9.1 | 6.5 | 6.1 | 0.50 | **0.025 / 0.015** | 10$^2$ |
| 11 | 10 | 3·10$^4$ | 0.02 | 15.6 | 13.5 | 6.1 | 5.1 | 0.89 | 0 / 0 | **2.5·10$^3$** |
| 12 | 10 | 3·10$^4$ | 0.02 | 11.9 | 7.7 | 4.4 | 3.6 | 8.5 | 1 / 0.61 | **2.5·10$^3$** |
| 13 | 10 | 3·10$^4$ | 0.02 | 15.6 | 15.0 | 5.6 | 4.5 | 0.83 | 0 / 0 | **2.5·10$^3$ [*3]** |
| 14 | 10 | 3·10$^4$ | 0.02 | 11.9 | 6.0 | 3.8 | 3.0 | 0.74 | 1 / 0.61 | **2.5·10$^3$ [*3]** |
|  |  |  |  |  | 8.6±2.0 [*1] | 10.5±4.4 [*2] | 7.4±4.7 [*2] | 0.47±0.1 |  |  |

**Table 1**. The list of simulation scenarios, with corresponding modeling parameters, and the calculated absorption. Columns from left to right: the number of the model run; the assumed values of stellar XUV and NUV fluxes at 1 a.u. in erg cm$^{-2}$ s$^{-1}$; helium abundance; the calculated mass loss rate of the planet in units of 10$^{10}$ g/s; the calculated total integrated absorption in the lines of HI (Lyα), OI, CII and HeI($2^3$S); the assumed equatorial surface value of magnetic field and the corresponding planetary magnetic dipole moment μ$_P$ in units of the Jovian magnetic moment μ$_J$ = 1.56 × 10$^{27}$ A m$^2$; the total mass loss rate of the star in units of 10$^{10}$ g/s, used for the simulation of SW.
The integration of absorption over the lines was performed in the following Doppler velocity intervals:
HI (Lyα): ±200 km/s, excluding ±50 km/s around the line center;
OI: ±50 km/s (averaged over the triplet) with the exclusion of ±6 km/s interval for the transition 1-2 at 1302 Å;
CII: ±90 km/s (averaged over the triplet) with the exclusion of ±6 km/s interval for the transition 3/2-1/2 at 1334.5 Å;
HeI: ±10 km/s around 10833.2 Å;
[*3] in these model runs a high-speed SW is simulated with velocity 300 km/s at the planet orbit.
The bottom row shows the measured absorption values analyzed in [*1] *Ben-Jaffel* (*2007, 2008*) and [*2] *Vidal-Madjar et al.* (*2004*)

Figure 1 shows the excess absorption at the position of metastable HeI($2^3$S) triplet at 10830 Å, obtained in the model runs N1, N2, N3, and N4 with different $F_{XUV}$ values. The corresponding He abundance values (see in Table 1) were chosen to fit approximately the measurement data. One can see a relatively good agreement of the simulations with the measurements of *Alonso et al.* (*2019*) in amplitude and line width. However, the model cannot reproduce the slight (~1.5 km/s) blue shift of the absorption maximum and the increased absorption in the blue wing of the line at the Doppler velocities range of [-20; -10] km/s. With a set of trial modelling runs we made sure that even a 5-fold reduction of $F_{NUV}$ cannot help in reproducing these features of the absorption profile. The line core is absorbed very close to the planet, at the altitudes of (1–1.5)$R_p$, where besides of photoionization, the reaction with H atoms also destroys efficiently the population of metastable HeI($2^3$S). Therefore, the acceleration time of HeI($2^3$S) atoms by the radiation pressure still remains short (in the regions where it matters).

Since the $F_{NUV}$ and $F_{XUV}$ values are not sufficiently constrained and could vary due to the natural fluctuations up to several times, the helium abundance iteratively derived by tuning of the modelling results to the measured HeI($2^3$S) absorption at 10830 Å can be constrained only by a factor of ~2.5.

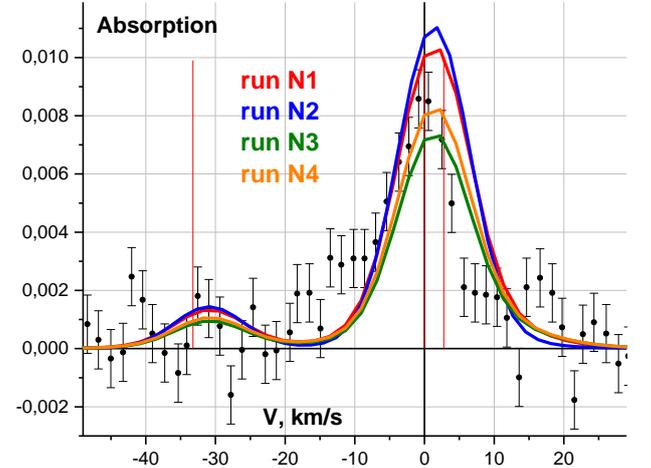

Figure 1. HeI($2^3$S) triplet absorption profiles, simulated with different $F_{XUV}$ and He abundances in the model runs N1, N2, N3, N4. Here and further on in similar plots the filled circles with error bars reproduce the measurements from *Alonso et al. (2019)* and red vertical lines indicate the positions of individual lines in the triplet.

### 3.2 Absorption at HI, OI, CII lines

To constrain further the major model parameters, we calculate the absorption in HI(Lyα) line and the resonant triplets of OI and CII, assuming the solar abundances for oxygen and carbon. As it was found out in our previous paper (*Shaikhislamov et al. 2020a*) and confirmed here, the increasing helium abundance decreases the absorption in all considered lines, because of the decrease of the atmospheric height scale (see, e.g., the model runs N2 and N5 for comparison). On the other hand, the absorption increases with the increasing $F_{XUV}$ because of the faster expansion and escape of the upper atmosphere of the planet. This trend is however reversed, because of the faster ionization of the absorbing elements (see, e.g., the model runs N3, N4, and N5 for comparison). Therefore, confining all the considered absorption profiles within the measurement error margins enables to restrict the ranges for the helium abundance and XUV flux at He/H=(2±0.25)·10$^{-2}$ and $F_{XUV}$=10±2 erg cm$^{-2}$ s$^{-1}$ (at 1 a.u.), respectively.

The best correspondence with observations is achieved in the model run N4. The worse fit among the considered lines takes place for the OI triplet, which shows relatively small absorption. It is necessary to mention, that in order to fit the depth of the absorption at HeI($2^3$S) triplet line addressed in the previous subsection, we took $F_{NUV}=3·10^4$ erg cm$^{-2}$ s$^{-1}$. The reduced $\chi^2$ for HeI($2^3$S) line in the range of Doppler velocities ±20 km/s in the model run N4 is 3.2. At the same time, simple shifting of the whole absorption profile by -1.5 km/s decreases the reduced $\chi^2$ down to 1.84. Thus, the question of whether the observed small (~1.5 km/s) blue shift is real remains important.

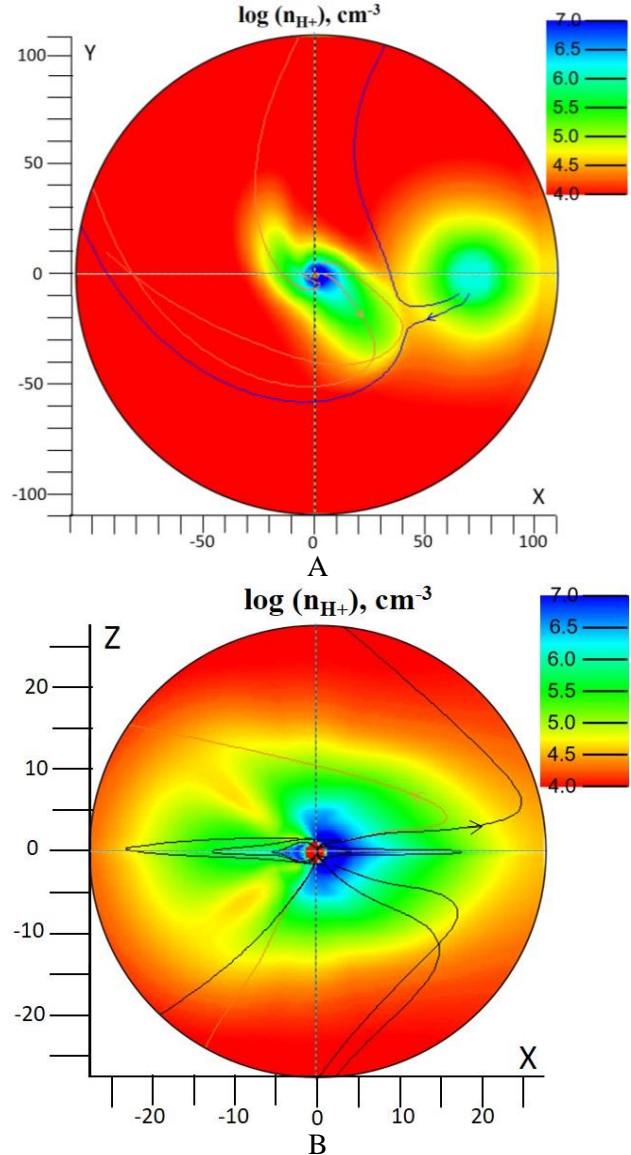

Figure 2. *Panel A:* proton density distribution in the orbital plane of the simulated domain, calculated in the model run N6, i.e., with $B_p$=1 G and other parameters the same as in the model run N4. The planet is at the center of the coordinate system (0,0) and moves anti-clockwise relative the star located at (70,0). *Panel B:* proton density distribution and MF lines (black) in the meridional plane around the planet. Proton fluid streamlines originated from the planet (orange) and from the star (blue) are shown in both panels. The axes, here and further on in the spatial distribution plots, are scaled in planetary radii $R_p$.

*3.3 Effect of the planetary magnetic field*

Imposing a relatively strong planetary dipole MF with a value of $B_p$=1 G at the surface of planet on the equator, which corresponds to the planetary magnetic dipole moment $\mu_P$ = 0,61 in units of the Jovian magnetic moment $\mu_J$ = 1.56 × 10$^{27}$ A m$^2$, profoundly changes the modelled atmospheric outflow of HD209458b, especially in the vicinity of the planet. This scenario is realized in the model run N6, which differs from the model run N4 just by the inclusion of the planetary MF. The color plot in Figure 2A shows the obtained distribution of the planetary and stellar proton density in the whole simulated domain, whereas Figure 2B presents the structure of the calculated planetary magnetosphere, viewed in the plane orthogonal to the ecliptic. As found out in our previous 2D simulations (*Khodachenko et al. 2015*) and predicted in earlier semi-analytic considerations (*Khodachenko et al. 2012*), the outflowing PW material stretches and opens the magnetic dipole field lines and forms an equatorial current layer, so-called magnetodisk. Figure 2 depicts also the material streamlines, showing that the flow of protons in general follows approximately the MF lines. We note however that this is only an approximate co-direction, because the neutral particles which do not feel MF exert an additional thermal pressure on the protons due to collisions, resulting in their diffusion across the field lines. The shape and structure of the formed magnetodisk-type electric current layer are shown in Figure 3. Although it surrounds the planet in the equatorial plane, it is not uniform.

The comparison of simulation results, obtained for the magnetized and non-magnetized planet, i.e. with $B_p$=1 G and $B_p$=0 G in the model runs N6 and N4, respectively, reveals first of all, the two times decrease of the planetary mass loss rate due to decreasing velocity and density of the escaping PW. Second, the MF causes significant reduction in the lines' absorption. To understand why this happens we provide in Figure 4 the density profiles of the corresponding absorbing elements. As can be seen, while the escaping PW material temperature remains almost the same, its velocity in the presence of MF is significantly reduced in a relatively extended (≈2.5$R_p$) region around the planet. This region corresponds to a so called 'dead' zone, where the planetary plasma escape is suppressed by the MF. Beyond the 'dead' zone, velocity sharply increases and quickly reaches the values typical for the non-magnetized case. Inside the thin magnetodisk, which has a thickness of about ±0.5 $R_p$, the MF is reduced, but it increases sharply in the lobes, whereas the plasma pressure behaves in the opposite way. Figure 4A shows the ratio of magnetic pressure at the edge of magnetodisk (at Z=0.5 $R_p$) to the plasma pressure at

Z=0. They appear in a close balance (if corrected for the lines curvature). This balance and the structure of magnetodisk have been addressed in *Khodachenko et al. (2015)*. It is worth to mention, that despite the strong influence of MF on the motion of escaping planetary material close to the planet, it does not influence the interaction between the PW and SW in a more distant region, because for the considered SW intensity the ionopause is located far beyond the magnetically dominated region.

observer's line of sight (LOS) at the altitude $r=2R_p$, (at the level of $Z=Y=1.4\ R_p$) i.e., beyond the photometric radius of the planet. One can see that when $B_p=1$ G, the densities of HI, OI, and CII are significantly reduced, especially on the night-side, behind the planet, where the influence of MF on the outflowing PW velocity, as compared to the thermal pressure, is the strongest. Detailed illustration of how the MF affects the absorption is given in Figure 5, where the spectral absorption profiles at the OI 1305 Å and CII 1336 Å lines are shown.

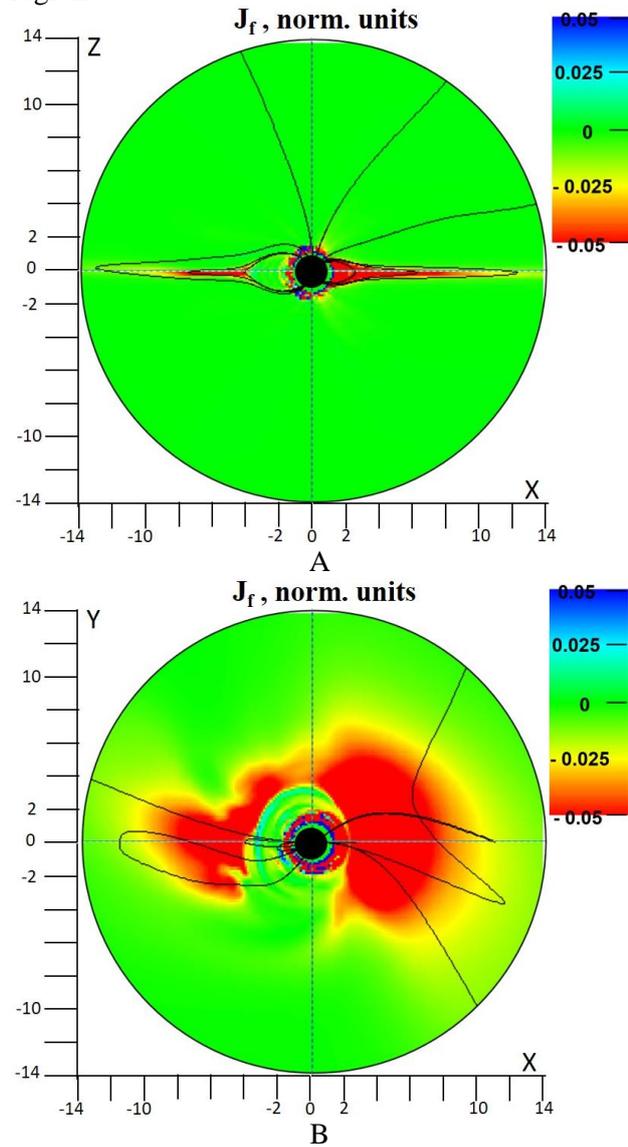

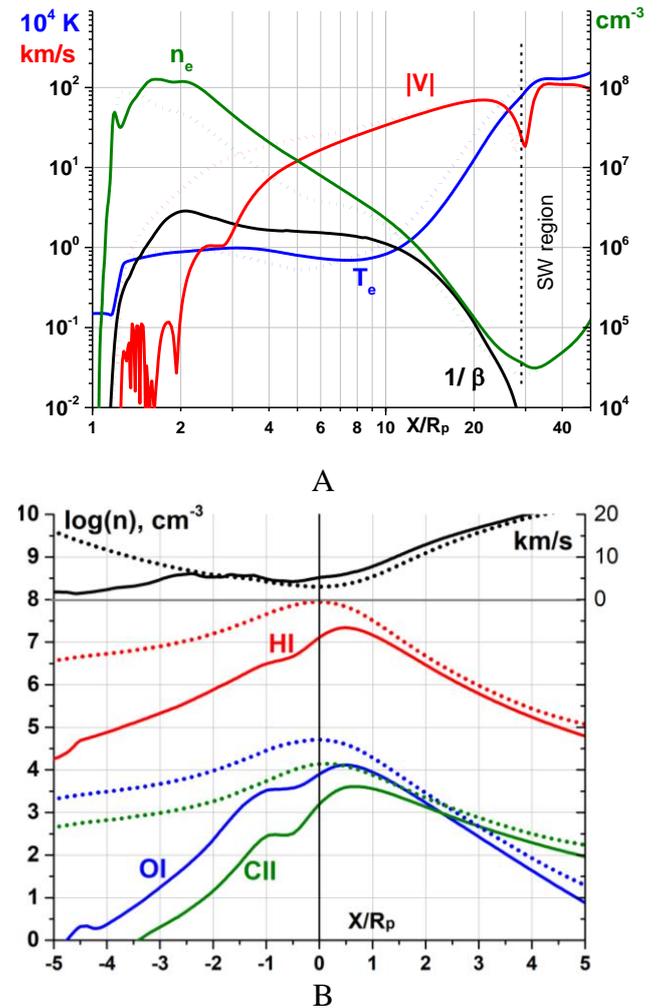

Figure 3. Distribution of the azimuthal component of electric current in the meridional (*Panel A*) and orbital (*Panel B*) planes, obtained in the model run N6. MF lines are shown in black.

The significantly reduced escape velocity in the case of a magnetized planet leads to a longer exposure of atoms to XUV radiation and consequently, to the stronger photo-ionization. Therefore, the reduced densities of HI, OI, and CII, clearly seen in panel B of Figure 4, would result in the decrease of absorption. The density profiles of HI, OI, and CII in Figure 4B are given along the

Figure 4. Distribution of major physical quantities along the observer's line of sight (X-axis), calculated for the magnetized ($B_p=1$ G, solid lines) and non-magnetized planet ($B_p=0$ G, dotted lines) in the model runs N6 and N4, respectively. *Panel A:* proton velocity (red, left axis); electron temperature (blue, left axis); electron density (olive, right axis); ratio of magnetic pressure at $Z=0.5\ R_p$ (the edge of magnetodisk) to plasma pressure at $Z=0$, i.e. analog of the inversed plasma-beta (black, right axis) along the planet-star line (at the level $Z=Y=0$). The vertical dashed line indicates the position of boundary between the planetary and SW plasmas (i.e., ionopause). *Panel B:* log of density of hydrogen atoms (red, left axis); neutral oxygen (blue, left axis); ionized carbon (olive, left axis); and flow velocity (black, right axis) at the level of $Z=Y=1.4\ R_p$.

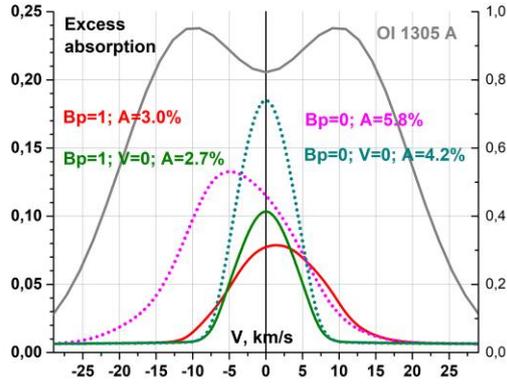

A

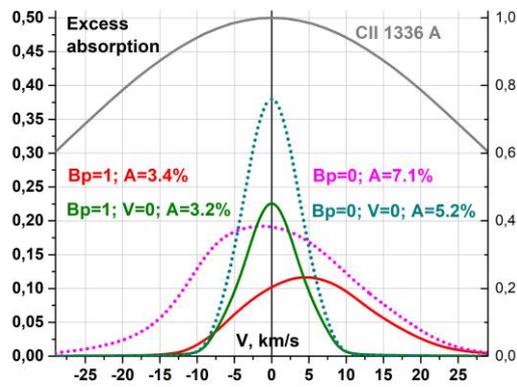

B

Figure 5. Absorption profiles at the OI 1305 Å (*Panel A*) and CII 1336 Å (*Panel B*) lines over Doppler-shifted velocity, calculated in the model runs N4 and N6. Dotted and solid lines correspond to the non-magnetized ($B_p=0$, model run N4) and magnetized ($B_p=1$ G, model run N6) cases, respectively; gray lines show the stellar emission line shapes used to calculate the absorption. Red and magenta lines show the full absorption in particular lines by the considered elements (to be compared with observations); olive and cyan lines show the full absorption calculated with the velocity of species artificially set to zero. The full integrated absorption values in each case are given in the legend.

Comparison of the spectral absorption profiles in Figure 5 reveals that the planetary MF results in narrowing of the absorption profile in the blue wing, due to the decreased velocity of the escaping PW, especially on the planet night-side; and the decreased absorption maxima in the line centers, due to the lower densities of absorbing elements around the planet (see, e.g., in Figure 4). The contribution of velocity V of the escaping planetary atmospheric material to the calculated absorption is demonstrated in Figure 5, which shows also the absorption profiles obtained with an 'ad hoc' prescribed zero velocity of PW. We found that for the magnetized planet the full integrated absorption values in the cases with V≠0 and V=0 are rather close. This confirms that the natural line broadening is a dominating mechanism of absorption in the magnetized HD209458b.

Without MF the velocity of escaping PW contributes about 1.5÷2 % to the absorption. This is due to the Doppler resonant line broadening, discussed in *Khodachenko et al. (2017)*. At the same time, the distribution of the absorbing particles across the LOS in front of the stellar disk reveals that in the case of magnetized planet they obscure the area of $\sim 3R_p$ around the planet, while in the case of non-magnetized planet this area extends up to $5R_p$.

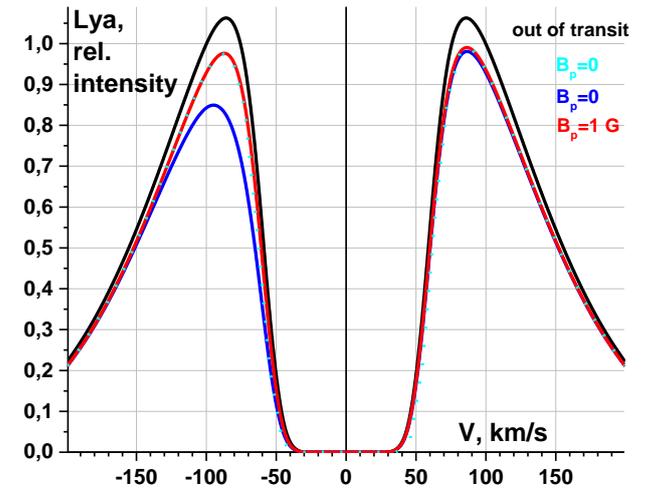

Figure 6. Simulated Lyα in-transit profiles versus the out-of-transit profile. *Cyan dotted line*: non-magnetized planet in moderate SW (model run N4); *blue line*: non-magnetized planet in strong SW (model run N11); *red line*: magnetized planet ($B_p=1$ G) in strong SW (model run N12); *black line*: the out-of-transit Lyα profile based on measurements by Vidal-Majar et al. (2003).

The calculation of absorption in the model runs N6-N10, for the decreasing values of $B_p$, from 1G down to 0.025G, with all other parameters of the system kept the same, reveals the gradual increase of the planetary mass loss and the absorption at all the considered lines, except of the HeI($2^3$S) line. By this, even for $B_p=0.1$ G its effect is still persistent, while being smaller than in the case of $B_p=1$ G. In particular, for $B_p=0.1$ G, the absorption at the lines of OI and CII appears 0.5% less than that in the non-magnetized case. Altogether, the closest match to the measurements in all considered lines is achieved in the model runs N9 and N10, which correspond to $B_p=$ (0.1-0.025) G, or the planetary magnetic dipole $\mu_P < 0.1$ in units of the Jovian value. The revealed stronger sensitivity of the absorptions at OI and CII lines to the value of MF, as compared with HI(Lyα) absorption, is due to their stronger relation with the moving PW controlled by the MF, whereas Lyα absorption on

HD209458b takes place in the relatively close to the planet regions within the Roche lobe with an insignificant material motion.

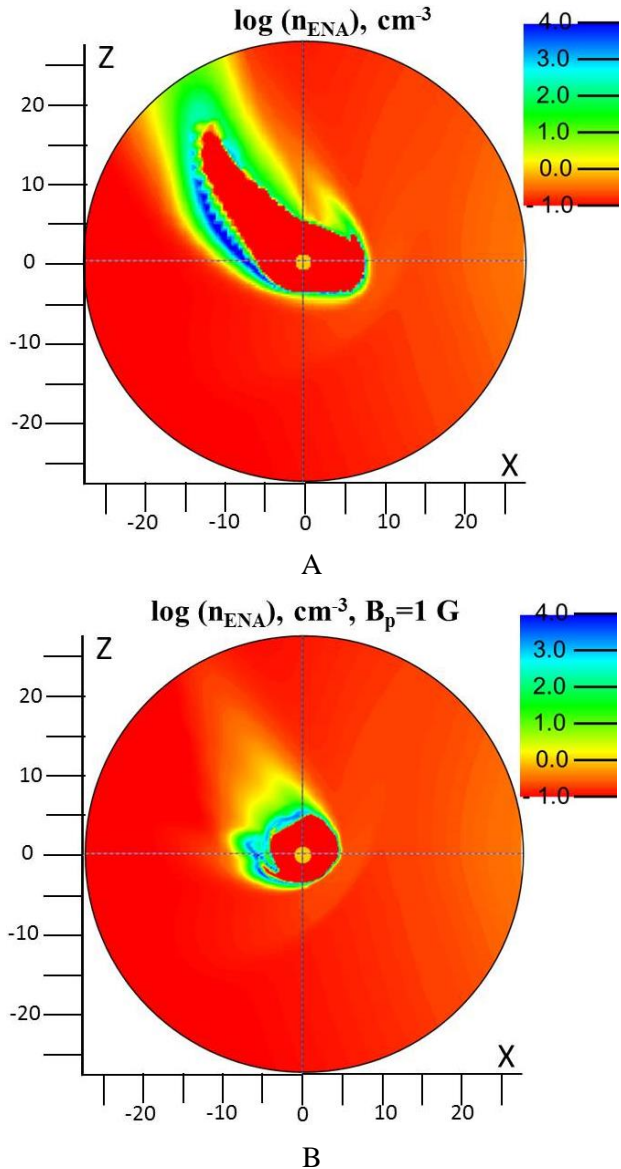

Figure 7. Distribution of ENAs in the orbital plane simulated under the conditions of strong SW in the cases of non-magnetized (*Panel A*, model run N11) and magnetized (*Panel B*, model run N12) planet.

*3.4 Magnetosphere under conditions of a strong SW*

*Khodachenko et al. (2017)* and *Shaikhislamov et al. (2020a)* have shown that a moderate SW does not influence the absorption at the considered lines, whereas the tenfold increased flux of the SW significantly increases the absorption in the blue wing of the Lyα. The latter happens due to the generation of significant amount of ENAs in the PW-SW interaction region. In this subsection we examine how planetary MF affects this scenario. First, in the model run N11 we simulated the case of a strong SW by taking the increased total stellar mass loss rate of $M^{/}_{SW}=2.5\times10^{13}$ g/s, with the increased up to $n_{sw}=10^5$ cm$^{-3}$ density at the planet orbit, while keeping the same as in the previously considered moderate SW cases velocity, $V_{sw}=220$ km/s and temperature $T_{sw}=8.5\cdot10^5$ K. The planetary MF was taken to be zero in this model run. Figure 6 shows the Lyα absorption profile which demonstrates under the conditions of the strong SW the absorption in the blue wing with average value of 19% in the range [-150; -50] km/s, which is much deeper than that in the case of a moderate SW. The enhanced absorption in the blue wing is also apparent for the OI and CII lines, though the total integrated absorption value is smaller.

It is worth to note, that for the magnetized planet, the Lyα absorption in the blue wing under the conditions of strong SW becomes practically the same as that in the case of a the non-magnetized planet and moderate SW (see in Figure 6). This is because the amount of ENAs in the magnetized case is suppressed by the MF, affecting the planetary and stellar plasma flows. It is demonstrated in Figure 7, which shows that in the case of the magnetized planet the population of ENAs becomes strongly depleted in the tail region. Altogether, the most of the absorption in the case of magnetized planet and strong SW, like in the case of non-magnetized planet and moderate SW, is due to the natural line broadening mechanism, and it is provided by the atomic hydrogen, confined within the magnetosphere and Roche lobe. This results in the symmetric Lyα absorption profiles obtained in the model runs N4 and N12 (Figure 6).

To study formation of the planetary magnetospheric obstacle and the role of the planetary MF, we consider further the case of a strong SW, but increase its velocity at the planet orbit up to 300 km/s, while keeping the total stellar mass loss rate, $M^{/}_{SW}=2.5\times10^{13}$ g/s, the same as in the model runs N11 and N12. This fast SW pushes the magnetopause closer to the planet and results in a similar structure of the magnetosphere as that known for the magnetized Solar System planets. The generated magnetospheric current system has the typical components, such as the dipolar MF dominated region around the planet, the magnetopause current sheet at the front of magnetosphere and the return current at high latitudes, as well as the tail current sheet, as seen in Figure 8A. However, the proton density distributions in Figure 8B show that the size of the PW dominated region around the planet in the case of $B_p$=1 G is practically the same as in the non-magnetized case. Therefore, we see that, even for a relatively high planetary MF, the modelled escaping PW of HD209458b is strong enough to dominate in the total pressure balance and to compete with the magnetic pressure at the distances of $(3 \div 4)R_p$.

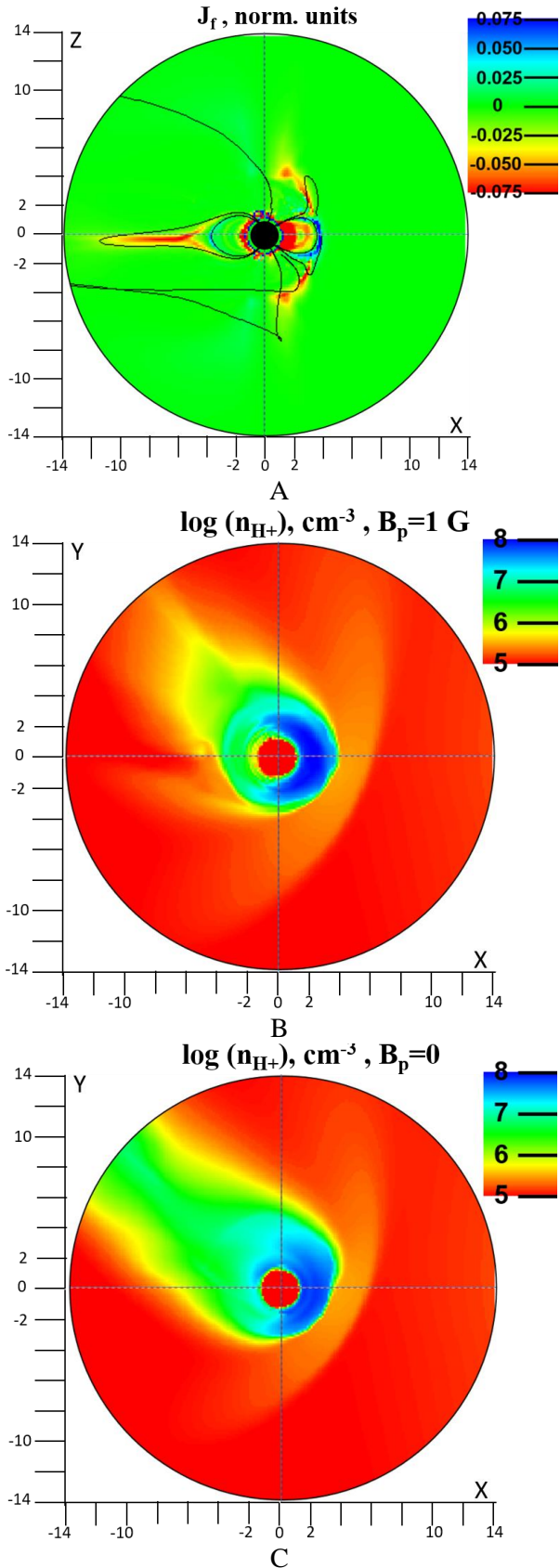

Figure 8. Magnetosphere structure of HD209458b and proton density distribution under the conditions of strong SW simulated in the model run N14. *Panel A:* Distribution of azimuthal current and MF lines in the meridional plane. *Panels B and C:* Proton density distribution in the orbital plane (*Panel B*) as compared to that in the case of non-magnetized planet (*Panel C*, model run N13).

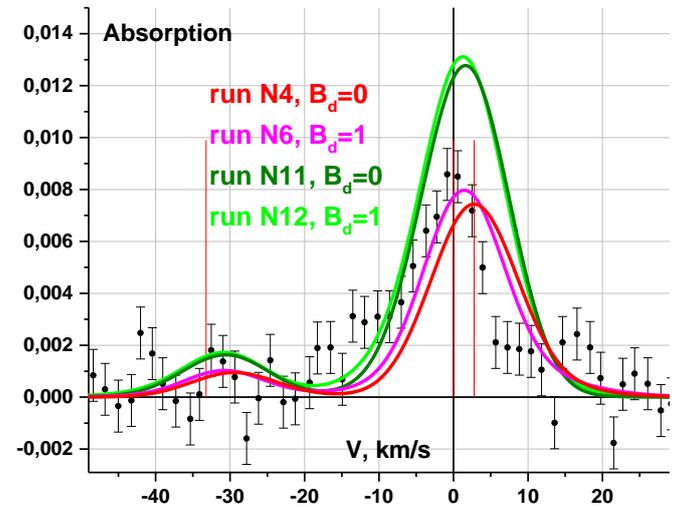

Figure 9. HeI($2^3$S) triplet absorption profiles, simulated with the moderate (model runs N4, N6) and strong (model runs N11, N12) SW in the cases of magnetized (model runs N6, N12) and non-magnetized (model runs N4, N11) planet.

Finally, we show in Figure 9 how the planetary MF and intensity of SW might affect the HeI($2^3$S) triplet absorption profiles. In particular, the overall effect of MF for $B_p$=1 G is rather small, though the profiles shift slightly towards the blue range. For the strong SW the absorption depth becomes $1.5 \div 2$ times higher. This is the consequence of a more compact magnetosphere and/or ionosphere compressed by the SW. At the same time, the blue shift of the HeI($2^3$S) absorption profile under the conditions of moderate SW and magnetized planet (the model run N6) provides better correspondence with the measurements in the blue part of the line as compared to the non-magnetized case (the model run N4), while in the red part the simulated absorption level is still higher than the observed one. At the same time, the simulated absorption in other considered lines still remains different from the measured values (see in Table 1).

## 4 DISCUSSION AND CONCLUSIONS

Using the global 3D multi-fluid HD and MHD models we simulated the measured HD209458b transit absorption depths at the far UV lines, and at the NIR line (10830 Å) of HeI($2^3$S) triplet. As continuation of our previous studies of HD209458b, the inclusion of the HeI($2^3$S) line into consideration and the comparison with corresponding measurements allows to constrain the helium abundance by He/H ~ 0.02, and the stellar XUV flux at 1 a.u. by $F_{XUV}$ ~10 erg cm$^2$ s$^{-1}$. While the simulated peak and half-width of the HeI($2^3$S) absorption profile fit the observations reasonably well, there is a net blue shift of ~1.5 km/s which the model cannot reproduce with the assumed variability of the parameters, affecting the HeI($2^3$S) line. At the same time, to verify the applied constrains on the helium abundance and the XUV flux values, and to exclude the effects of the stellar short-periodic variability, more measurements are needed. For example, two available measurements at 10830 Å for the HD189733b show a factor of 1.5 difference, which could be related with the varying stellar radiation.

For the first time, we study in this paper the influence of the planetary dipole MF on the absorption in the most of spectral lines simultaneously, for which the measurements have been made. Our MHD simulations show that the planetary magnetic dipole moment $\mu_P$ = 0.61 (in units of the Jovian magnetic moment), which provides the MF equatorial surface value of 1 G, profoundly changes the character of the escaping PW flow and the related absorption. The total mass loss rate in this case is reduced by 2 times, as compared to the non-magnetized planet. We note that our previous 2D MHD modelling with similar MF as well as and other parameters as those used here, but without the account of the incoming SW plasma flow, resulted in overestimation of the mass loss reduction, but the physical nature of the effect remains the same. It is related with the suppression of the planetary material escape by the MF and confinement of the part of the escaping plasma within the dead-zone inside the planetary magnetosphere. All the basic phenomena related to the influence of the planetary dipole MF on the structure of the escaping PW, and revealed in the 2D MHD modelling of the magnetized analog of HD209458b (*Khodachenko et al. 2015*), were also reproduced in the present 3D MHD simulations. In particular we see the formation of the 'dead'- and 'wind'- zones around the planet with their different character of plasma motion. The 3D MHD modelling also confirmed the result of previous 2D simulations that the escaping PW forms a thin magnetodisk in the equatorial region around the planet. The significantly reduced velocity of PW at the low altitudes around the planet, and especially at the night side, results in the stronger photo-ionization of species and significantly lower densities of the corresponding absorbing elements. Altogether, these reduced velocities and lower densities result in significant decrease of the absorption at Lyα (HI), OI, and CII lines, though the absorption at HeI($2^3$S) line remains nearly the same.

As it was shown in our previous papers, the dense and fast SW interacting with the escaping upper atmosphere of HD209458b, generates sufficient amount of ENAs to produce significant absorption in the high-velocity blue wing of the Lyα line. However, according to the reported here 3D MHD modelling, the planetary magnetic dipole moment of 0.61$\mu_J$ with the equatorial MF surface value $B_p$=1 G prevents the formation of ENAs, especially at the night side. This effect opens a possibility to constrain the range of planetary MF values for the evaporating hot Jupiters and warm Neptunes subject to strong enough SW. As an example of such cases, may be Pi Men C, which reveals rather low Lyα absorption (*García Muñoz et al. 2020, Shaikhislamov et al. 2020b*), despite of qualitative similarity to other warm Neptunes, e.g., GJ436b and GJ3470b, where significant blue shifted Lyα absorption generated by ENAs was measured (*Ehrenreich et al. 2015, Lavie et al. 2017, Bourrier et al. 2018*) and simulated (*Khodachenko et al. 2019, Shaikhislamov et al. 2020c*). As to the absorption at the HeI($2^3$S) line, the strong SW also results in its increase, though the influence of MF in this case is negligible.

At the same time, the performed modelling has shown that all the considered spectral lines are sensitive, while differently, to the presence and value of the MF. The most sensitive in that respect are the lines of HI(Lyα), OI, CII, but also the absorption at HeI($2^3$S) line has shown certain weak dependence on the MF. This is related with the strong connection of the absorption with the density and dynamics of the absorbing particles which are controlled by the MF, affecting the global dynamics of the planetary material flow, interacting with the SW. At the same time, other free parameters of the system, such as elements abundances, stellar radiation fluxes, and SW intensity, besides of the MF, also affect the observed spectral features, and should be constrained. This altogether makes the detection and diagnostics of the planetary MF on the basis of the spectral absorption profiles to be a complex multi-parametric problem. It has to involve the whole range of the available spectral measurements, which should be fitted with the model simulations.

The simulation results presented in this paper indicate that the magnetic dipole moment $\mu_P$ of HD209458b should be at least an order of magnitude less than that of the Jupiter. This conclusion agrees with the previous estimates, based on more simplified models (e.g., *Kislyakova et al. 2014*) and less observational data (only Lyα absorption was considered). The application of 3D MHD models, like e.g., that presented here, which self-consistently simulate the escape of upper atmospheres of hot exoplanets and the related transits at several spectral lines available for measurement and sensitive to the PW dynamics affected by the MF, opens the way for probing and quantifying exoplanetary MFs by fitting the modelling results to observations. This altogether will shed more light on the nature of exoplanetary magnetism and related magnetospheric phenomena.

Acknowledgements:
This work was supported by grant № 18-12-00080 of the Russian Science Foundation and grant No.075-15-2020-780 (GA No.13.1902.21.0039) of the Russian Ministry of Education and Science. MLK acknowledges the projects I2939-N27 and S11606-N16 of the Austrian Science Fund (FWF). HL acknowledges the projects P25256-N27 and S11607-N16 of the Austrian Science Foundation (FWF). IS is thankful to the Programme "Astrophysical Origins: Pathways from Star Formation to Habitable Planets" (July 2019, ESI, Wien, Austria) for productive discussions. Parallel computing simulations, key for this study, have been performed at Computation Center of Novosibirsk State University, SB RAS Siberian Supercomputer Center and Joint Supercomputer Center of RAS.

Data availability: All the underlying data were published in the cited papers. Details of the modelling code may be shared on a reasonable request.